\documentstyle[epsfig]{article}
\begin{document}
\thispagestyle{empty}
\setcounter{page}{0}
\begin{flushright}
UWThPh-1998-20\\
April 1998
\end{flushright}
\vfill
\begin{center}

%EndAName
{\LARGE \bf Bohm's Interpretation of Quantum Mechanics and the
Reconstruction of\\[10pt] the Probability Distribution}\\[40pt]
\large
Heinz Rupertsberger\\[0.5cm]
Institut f\"{u}r Theoretische Physik, Universit\"{a}t Wien\\
Boltzmanngasse 5, A-1090 Wien, Austria \\[10pt]
\end{center}
\vfill

%\date{}
%\maketitle

\begin{abstract}
Within Bohm's interpretation of quantum mechanics particles follow
``classical'' trajectories that are determined by the full solution of the
time dependent Schr\"{o}dinger equation. If this interpretation is
consistent it must be possible to determine the probability
distribution at time $t$, $\rho (x,t)$, from the probability distribution at
time $t=0$, $\rho (x,0)$, by using these trajectories. In this paper it is
shown that this is the case indeed.
\end{abstract}

\newpage

\section{Introduction}

In 1952 David Bohm \cite{Bohm52} proposed a new interpretation of quantum
mechanics connecting ``classical'' trajectories to particles with a
space-time probability distribution given by the full solution of the time
dependent Schr\"{o}dinger equation. A detailed presentation of this
interpretation may be found for instance in \cite{Bohm93}, whereas its
historical contingency is discussed in \cite{Cushing94}.

The main points of his interpretation, which are of concern here, may be
described as follows (for simplicity the space dimension is taken to be $1$
and units $\hbar =m=1$ are used). Let $\psi (x,t)$ be the solution of the
time dependent Schr\"{o}dinger equation for a certain system, that is 
\begin{equation}
i\frac{\partial }{\partial t}\psi (x,t)=H\psi (x,t)
\end{equation}
where $H$ denotes the Hamilton operator of the system. Then one decomposes $%
\psi (x,t)$ in its absolute value and the corresponding phase which results
in 
\begin{equation}
\psi (x,t)=R(x,t)\exp\left[iS(x,t)\right]
\end{equation}
The phase $S(x,t)$ determines the ``classical'' motion by requiring that the
momentum of a particle at position $x$ at time $t$ is given by 
\begin{equation}
p=\frac{dx}{dt}=\frac{\partial S(x,t)}{\partial x} \label{p-def}
\end{equation}
Solving (\ref{p-def}) for a particle with initial position $x_{0}$ at time $%
t_{0}$ attributes to this particle a classical trajectory $x(x_{0},t)$
subject to the initial condition $x_{0}=x(x_{0},t_{0})$. Obviously one may
use these trajectories to transport the probability distribution 
\begin{equation}
\rho (x,t_{0})=\left| \psi (x,t_{0})\right| ^{2}=R(x,t_{0})^{2}
\end{equation}
in time by moving the initial distribution at points $x_{0}$ to the points $%
x(x_{0},t)$, which is suggested by Bohm's interpretation of quantum
mechanics. Of course this transport in time cannot be the complete procedure
since the normalization condition of the probability distribution must be
maintained too. In the next section it will be shown how one can reconstruct
the probability distribution $\rho (x,t)$ from the initial one, $\rho
(x,t_{0})$, by using classical trajectories, the solutions of (\ref
{p-def}), and that the result agrees with the usual quantum mechanical one,
namely 
\begin{equation}
\rho (x,t)=\left| \psi (x,t)\right| ^{2}  \label{usual_quant}
\end{equation}

\section{Reconstruction of the Probability Distribution}

Let the probability distribution at time $t=0$ be  ($t_{0}$ is chosen to be $0$
for simplicity)
\begin{equation}
\widetilde{\rho }(x,0)=\int \delta (x-x_{0})f(x_{0})dx_{0}=f(x)
\end{equation}
This is the continuous analogue of particles distributed at points $x_{0}$
with some probability $f(x_{0})$. According to Bohm's interpretation the
points $x_{0}$ move to the points $x(x_{0},t)$ in time and therefore
following this idea the probability distribution at time t must be given by 
\begin{equation}
\widetilde{\rho }(x,t)=\int \delta ((x-x(x_{0},t))f(x_{0})dx_{0}
\label{rho_time}
\end{equation}
where $x(x_{0},t)$ is the solution of (\ref{p-def}) with the appropriate
initial condition. From (\ref{rho_time}) it is easy to obtain the wanted
result because the integration can be done without difficulty by means of
the $\delta $-function. The zero of the $\delta $-function is given by $%
x_{0}(x,t)$, the inverse function to $x(x_{0},t)$. This gives, using
differentiation rules for inverse functions, 
\begin{equation}
\widetilde{\rho }(x,t)=\left| \frac{\partial x_{0}(x,t)}{\partial x}\right|
f(x_{0}(x,t))
\end{equation}
and the final result is 
\begin{equation}
\widetilde{\rho }(x,t)=\left| \frac{\partial x_{0}(x,t)}{\partial x}\right| 
\widetilde{\rho }\left( x_{0}\left( x,t\right) ,0\right)  \label{rho_final}
\end{equation}
That this result, obtained by using Bohm's interpretation of quantum
mechanics, agrees with the usual quantum mechanical result (\ref{usual_quant}%
) can be shown as follows. Together with the current density 
\begin{equation}
j(x,t)=\int \frac{\partial x(x_{0},t)}{\partial t}\delta
(x-x(x_{0},t))f(x_{0},t)dx_{0}
\end{equation}
$\widetilde{\rho }(x,t)$ as defined in (\ref{rho_time}) fulfills the
continuity equation 
\begin{equation}
\frac{\partial \widetilde{\rho }(x,t)}{\partial t}+\frac{\partial j(x,t)}{%
\partial x}=0
\end{equation}
But $j(x,t)$, due to the momentum definition (\ref{p-def}), can also be
written in the form 
\begin{equation}
j(x,t)=\int \frac{\partial S(x,t)}{\partial x}\delta
(x-x(x_{0},t))f(x_{0},t)dx_{0}=\frac{\partial S(x,t)}{\partial x}\widetilde{%
\rho }(x,t)
\end{equation}
Thus $\widetilde{\rho }(x,t)$ fulfills the same continuity
equation as $\rho (x,t)$ defined in (\ref{usual_quant}).
At time $t=0$ both agree, if $f(x)$ is chosen to be
\begin{equation}
f(x)=\left| \psi (x,0)\right| ^{2}
\end{equation}
The uniqueness of the solution of the continuity equation implies then that 
\begin{equation}
\widetilde{\rho }(x,t)\equiv \rho (x,t)\quad if\quad \widetilde{\rho }%
(x,0)=\rho (x,0)
\end{equation}
This shows that the reconstruction of the probability distribution at time $%
t=0$ following Bohm's ideas agrees with the usual quantum mechanical result
and the explicit expression is given by (\ref{rho_final}). But the
imagination that the probability distribution at time $t$ ist just obtained
by the flow of particles from the initial distribution following the
trajectories $x(x_{0},t)$ is in general wrong due to the factor $\left| 
\frac{\partial x_{0}(x,t)}{\partial x}\right| $ in (\ref{rho_final}). This
factor could also be interpreted as a normalization factor since due to it
the normalization of the wave function is valid for all times 
\begin{equation}
1=\int \left| \psi \left( x_{0},0\right) \right| ^{2}dx_{0}=\int \left| \psi
\left( x_{0}\left( x,t\right) ,0\right) \right| ^{2}\left| \frac{%
dx_{0}\left( x,t\right) }{dx}\right| dx
\end{equation}
The next section will show the role of this factor within three
different examples explicitly.

\section{Examples}

\subsection{The Coherent State of the Harmonic Oscillator}

The normalized coherent state with amplitude $d$ is given by ($\omega =1$) 
\begin{equation}
\psi _{d}(x,t)=\frac{1}{\pi ^{\frac{1}{4}}}\exp \left\{ -i\left[ \frac{t}{2}%
-\frac{d^{2}}{4}\sin 2t+dx\sin t\right] -\frac{1}{2}\left[ x-d\cos
t\right] ^{2}\right\} 
\end{equation}
Which gives for the absolute value $R(x,t)$ and the phase factor $S(x,t)$%
\begin{eqnarray}
R\left( x,t\right)  &=&\frac{1}{\pi ^{\frac{1}{4}}}\exp \left[ -\frac{1}{2}%
\left( x-d\cos t\right) ^{2}\right]  \\
S(x,t) &=&-\frac{t}{2}+\frac{d^{2}}{4}\sin 2t-dx\sin t
\end{eqnarray}
For the definition of the momentum one obtains therefore
\begin{equation}
p=\frac{\partial S}{\partial x}=-d\sin t
\end{equation}
resulting in the equation of motion 
\begin{equation}
\frac{dx}{dt}=-d\sin t
\end{equation}
with the solution 
\begin{equation}
x(x_{0},t)=d\left( \cos t-1\right) +x_{0}
\end{equation}
From this the inverse function is given by 
\begin{equation}
x_{0}(x,t)=x-d\left( \cos t-1\right) 
\end{equation}
and therefore 
\begin{equation}
\left| \frac{\partial x_{0}(x,t)}{\partial x}\right| =1
\end{equation}
as expected since a coherent state does not change its shape in time.
Obviously the reconstructed probability distribution agrees with the quantum
mechanical result as it has to be 
\begin{eqnarray}
\rho (x,0) &=&R(x,0)^{2}=\frac{1}{\pi ^{\frac{1}{2}}}\exp \left[ -\left(
x-d\right) ^{2} \right] \Longrightarrow  \\
\rho (x_{0}(x,t),0) &=&\frac{1}{\pi ^{\frac{1}{2}}}\exp
\left\{-\left[x_{0}(x,t)-d\right]^{2}\right\}=R(x,t)^{2}  \nonumber
\end{eqnarray}

\subsection{The Wave Packet for a Moving Free Particle}

The normalized wave function for a free particle with a Gaussian
distribution around $x=0$ at time $t=0$ and for which the peak of the
probability distribution moves with constant velocity (here taken to be
one) in time is given by
\begin{equation}
\psi (x,t)=\frac{1}{\pi ^{\frac{1}{4}}\sqrt{1+it}}\exp \left[ -\frac{%
x^{2}-2ix+it}{2\left( 1+it\right) }\right] 
\end{equation}
From this the absolute value and the phase of the wave function are given by
\begin{equation}
R(x,t)=\frac{1}{\sqrt[4]{\left( t^{2}+1\right) \pi }}\exp \left[-\frac{1}{2}\frac{%
\left( t-x\right) ^{2}}{t^{2}+1}\right]
\end{equation}
\begin{equation}
S(x,t)=\frac{1}{4i}\ln \frac{1-it}{1+it}+\frac{tx^{2}+2x-t}{2(1+t^{2})}
\end{equation}
with the corresponding ``classical'' momentum
\begin{equation}
p=\frac{\partial S}{\partial x}=\frac{tx+1}{1+t^{2}}
\end{equation}
The equation of motion
\begin{equation}
\frac{dx}{dt}=\frac{tx+1}{1+t^{2}}
\end{equation}
has the solution
\begin{equation}
x=t+x_{0}\sqrt{1+t^{2}}
\end{equation}
whose inverse is given by
\begin{equation}
x_{0}=\frac{x-t}{\sqrt{1+t^{2}}}
\end{equation}
and
\begin{equation}
\left| \frac{\partial x_{0}(x,t)}{\partial x}\right| =\frac{1}{\sqrt{1+t^{2}}%
}
\end{equation}
which is just the right factor to correct the normalization for the
broadening of the wave packet in time. Again one obtains complete agreement
with Bohm's interpretation
\begin{eqnarray}
\rho (x,0) &=&R(x,0)^{2}=\frac{1}{\sqrt{\pi }}\exp (-x^{2})\Longrightarrow
\left| \frac{\partial x_{0}(x,t)}{\partial x}\right| \rho (x_{0}(x,t),0) \\
&=&\frac{1}{\sqrt{(1+t^{2})\pi }}\exp\left[-\frac{(x-t)^{2}}{1+t^{2}}%
\right]=R(x,t)^{2}  \nonumber
\end{eqnarray}
One may be tempted to interpret the factor $\left| \frac{\partial x_{0}(x,t)%
}{\partial x}\right| $ to be responsible only for the normalization of the
wave function without any relation to its shape. That this imagination is
completely wrong will be shown in the last example.

\subsection{The Harmonic Oscillator Again}

Instead of a coherent state a superposition of the ground state with the
first excited state of the harmonic oscillator is considered. The time
dependent solution for this case is given by
\begin{equation}
\psi (x,t)=\sqrt{\frac{2}{3}}\frac{1}{\pi ^{1/4}}\exp \left( -\frac{x^{2}+it%
}{2}\right) \left[ 1+x\exp \left( -it\right) \right] 
\end{equation}
and therefore
\begin{equation}
R(x,t)=\sqrt{\frac{2}{3}}\frac{1}{\pi ^{1/4}}\exp \left( -\frac{x^{2}}{2}%
\right) \sqrt{1+2x\cos t+x^{2}}
\end{equation}
\begin{equation}
S(x,t)=-\frac{t}{2}+\frac{1}{2i}\ln \frac{1+x\exp \left( -it\right) }{%
1+x\exp (it)}
\end{equation}
Proceeding in the usual way one has therefore to solve the differential
equation
\begin{equation}
\frac{dx}{dt}=-\frac{\sin t}{1+2x\cos t+x^{2}}
\end{equation}
which has the implicit solution
\begin{eqnarray}
&&\frac{3}{4}\sqrt{\pi }\textrm{erf}\left( x\right) -\exp \left(
-x^{2}\right)
\cos t-\frac{1}{2}x\exp \left( -x^{2}\right)  \\
&=&\frac{3}{4}\sqrt{\pi }\textrm{erf}\left( x_{0}\right) -\exp \left(
-x_{0}^{2}\right) \cos t-\frac{1}{2}x_{0}\exp \left( -x_{0}^{2}\right)  
\nonumber
\end{eqnarray}
One can numerically extract from this implicit solution $x(x_{0},t)$ or the
inverse $x_{0}(x,t)$ and compare the exact result $\rho (x,t)$ with the
result obtained from using ``classical'' trajectories but without the
factor $\left| \frac{\partial x_{0}(x,t)}{\partial x}\right| $, that is with $%
\rho (x_{0}(x,t),0)$ only. This is shown in Fig. 1 for the time $t=\pi$ (for simplicity the square roots are plotted, that is $R(x,t)$ and $%
R(x_{0}(x,t),0)$). The disagreement in shape is obvious, but full agreement is achieved if the factor $\left| \frac{\partial
x_{0}(x,t)}{\partial x}\right| $ is taken into account.
\begin{figure}[t]
\begin{center}
\epsffile{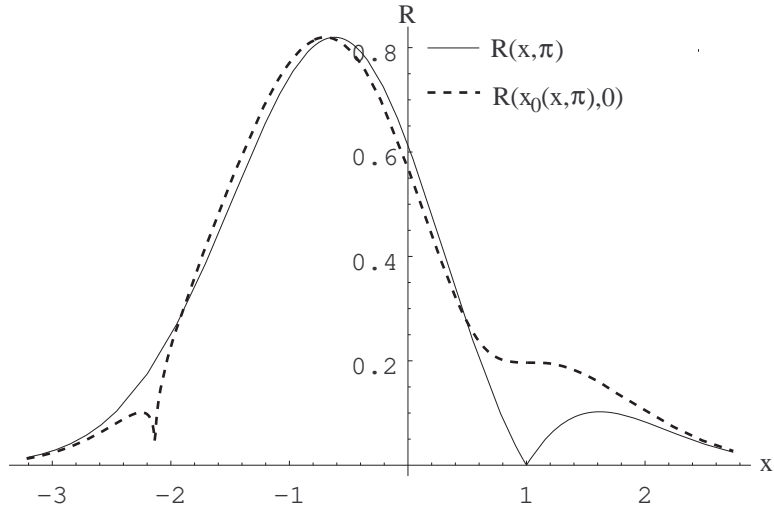}
\caption{Comparison of the exact result $R(x,\pi)$ and the result obtained by
using ``classical'' trajectories but without the ``normalization''
factor}
\end{center}
\end{figure}  
\section{Conclusion}
It has been shown that Bohm's interpretation of quantum mechanics gives the
correct answer for the probability distribution at later times when applied
to the probability distribution of a system at time $t=t_{0}$. Of course no
new results can be obtained in this way, since first one has to know the
wave function of the system for all times. But the main point is that the
interpretation in itself does not lead to any contradictions if one takes it
literally. Nevertheless one has to be very cautious in applying this
interpretation. As is shown by the last example of the superposition of two
harmonic oscillator eigenstates, one cannot view the particle flow as just
transferring the initial proabability distribution to a final one, which has
to be normalized only, but agrees in shape with the result obtained in this
way. On the contrary, the factor $\left| \frac{\partial x_{0}(x,t)}{\partial x%
}\right| $ may change this shape quite drastically in general and makes it
very difficult to visualize what is going on.


\begin{thebibliography}{9}
\bibitem{Bohm52}  D. Bohm, Phys.\ Rev. 85 (1952) 166, 180.

\bibitem{Bohm93}  D. Bohm, B.J. Hiley, ``The Undivided Universe: An
Ontological Interpretation of Quantum Theory'', Routledge, London, 1993.

\bibitem{Cushing94}  J.T. Cushing, ``Quantum Mechanics: Historical
Contingency and the Copenhagen Hegemony'', University of Chicago Press,
Chicago, 1994
\end{thebibliography}
\end{document}